\documentclass[10pt]{ismd08}
\usepackage{graphicx}
\usepackage{amssymb}

\begin{document}

\title{Central Exclusive Production at the Tevatron}

\author{Michael G. Albrow \\
 on behalf of the CDF Collaboration.}
\institute{Fermi National Accelerator Laboratory,\\ 
P.O.Box 500, Wilson Road, Batavia, IL 60510, USA
}
\maketitle

\begin{abstract}
 In CDF we have observed several exclusive processes:
 $\gamma\gamma\rightarrow e^+e^-$ and $\mu^+\mu^-$, $\gamma + I\!\!P \rightarrow J/\psi, \psi(2S)$, and
 $I\!\!P+I\!\!P \rightarrow \chi_c$. The cross sections agree with QED, HERA photoproduction data, and
 theoretical estimates of $gg \rightarrow \chi_c$ with another gluon exchanged to screen the color. This 
 observation of exclusive $\chi_c$, together with earlier observations of exclusive dijets and exclusive
 $\gamma\gamma$ candidates, support some theoretical predictions for $p+p\rightarrow p+H+p$ at the LHC.
 Exclusive dileptons offer the best means of precisely calibrating forward proton spectrometers. 
\end{abstract}


\section{Central Exclusive Production}
  Central exclusive production at the Tevatron is the process $p + \bar{p} \rightarrow p + X + \bar{p}$, 
  where ``$+$'' means a
  rapidity gap $\Delta y$ exceeding 3 units, and $X$ is a simple system fully measured. 
  Exchanges ($t$-channel) over such large gaps must be color singlets with spin $J$ [or Regge intercept $\alpha(0)$]
  $\geq$ 1.0. Only photons $\gamma$ and pomerons $I\!\!P$ qualify, apart from $W$ and $Z$ bosons
 which always cause the proton to break up. The gluon $g$ would qualify apart from its color, but if
 another gluon is exchanged that can be cancelled, and $I\!\!P = gg$ is often a good
 approximation. It cannot be exact; QCD forbids a pure $gg$ state, and a $q\bar{q}$ component certainly
 grows as $Q^2$ increases. The $I\!\!P$ has C = +1; in QCD one should also have a $ggg$ state with C = -1, the
 odderon~\cite{ewerz}
 $O$, not yet observed. The central
  masses $M_X$ are roughly limited to $M_X \lesssim \frac{\sqrt{s}}{20}$ with the outgoing protons having Feynman
  $x_F >$ 0.95. Hence $M_X \lesssim $ 3 GeV at the CERN ISR~\cite{afs}, appropriate for glueball spectroscopy,
  where $M(\pi^+\pi^-)$ shows a broad $f_0(600)$, a narrow $f_0(980)$ and still unexplained structure possibly
  associated with $f_0(1710)$, a glueball candidate. The study of $X$ = hadrons, e.g. $\phi\phi$ and $D^\circ \bar{D}^\circ$ to
  name two channels among many, has not been studied above ISR energies, but CDF is a perfect place to do
  it and hopefully it will be done~\cite{kiev}.
  
  At the LHC $M_X$ can reach $\approx$ 700 GeV, into the electroweak sector,
  and we can have $X = Z, H, W^+W^-, ZZ $, slepton pairs $\tilde{l}\tilde{l}$, etc. Measuring the forward protons after 120m of 8T
  dipoles, in association with the central event, as the FP420~\cite{fp420} proponents hope to do at
  ATLAS and CMS, one can measure $M_X$ with $\sigma(M_X) \approx$ 2 GeV per event~\cite{albrow}, and for a state
  such as $H$, also its width if $\Gamma(H) \gtrsim$ 3 GeV/c$^2$.  There are scenarios (e.g. SUSY) in which FP420 could provide
  unique measurements, e.g. if there are two nearby states both decaying to $b\bar{b}$ or to $W^+W^-$. 
  The quantum numbers of $X$ are $J^{PC} = 0^{++}$ or $2^{++}$ (and these are distinguishable) for $I\!\!P I\!\!P$
  production. 
  Two-photon collisions $\gamma\gamma\rightarrow l^+l^-, W^+W^-, \tilde{l}\tilde{l}$ 
  become important at the LHC thanks to the intense high momentum 
  photons, orders of magnitude more than at the Tevatron, giving $>50$ fb for $W^+W^-$ as a continuum background to $H\rightarrow W^+W^-$.
  $H\rightarrow ZZ$ does not have this background.
  
  While there is a gold mine of physics in $p+X+p$ at the LHC, we need to show
  that (a) the cross sections are within reach, and (b)
  one can build the spectrometers with resolution $\sigma(M_X)\approx$ 2 GeV/c$^2$ and calibrate
  their momentum scale \emph{and resolution}, to measure
  $\Gamma(H)$, and perhaps to distinguish nearby states. Both these issues are addressed
  by CDF in a ``TeV4LHC'' spirit, and they are also very interesting in their own right.  The calculation
  of cross sections (e.g.~\cite{khoze}) involves, in addition to $\sigma(gg \rightarrow X)$, the unintegrated gluon
  distribution $g(x_1, x_2)$, rapidity gap survival probability (no other parton interactions), and the
  Sudakov factor (probability of no gluon radiation producing hadrons). The Durham group predicts $\sigma(SMH)$
  for $p+H+p$ at the LHC = 3$^{\times 3}_{\div 3}$ fb. At the Tevatron $p+H+\bar{p}$ is out of reach, but the
  process $p+ \chi_c(\chi_b) + \bar{p}$ is identical as far as QCD is concerned, as is $ p + \gamma\gamma +\bar{p}$. 
  Measuring these constrains the $SMH$ cross section.
  In CDF we have looked for both exclusive $\gamma\gamma$ \cite{exclgg} and $\chi_c$ \cite{mumu34}, without however having detectors able to 
  see the $p$ and $\bar{p}$. Instead we added forward calorimeters ($3.5 < |\eta| < 5.1$) and beam shower
  counters BSC ($5.5 < |\eta| < 7.4$). If these are all empty there is a high probability
  that both $p$ and $\bar{p}$ escaped intact with small $|t|$. We also measured~\cite{goulianos} exclusive
  dijets. 
     
     For the exclusive $\gamma\gamma$ search we triggered on events with two electromagnetic ($EM$) clusters with $E_T >$ 4
     GeV in the central
     calorimeter, with a veto on signals in the BSC. This killed pile-up events and
     enabled us to take data without prescaling the trigger. We required 
     all other detectors to be consistent with only noise; then our \emph{effective} luminosity is 
     only about 10\% of the delivered luminosity. We found~\cite{exclgg} 3 events with exactly two back-to-back 
     $EM$-showers (assumed to be photons) with
     $M(\gamma\gamma)>$ 10 GeV/c$^2$. From wire proportional chambers at the shower maximum we concluded that two were perfect
     $p + \bar{p} \rightarrow p + \gamma\gamma+\bar{p}$ candidates and one was also consistent with being a
     $p + \bar{p} \rightarrow p +\pi^\circ\pi^\circ+\bar{p}$ event. The
     Durham prediction~\cite{durhgg} was 0.8$^{\times 3}_{\div 3}$ events, clearly consistent. We have since accumulated
     more data, with a lower threshold, now being analysed.

     With the above trigger we also found~\cite{exclee} 16 $p + \bar{p} \rightarrow p + e^+e^- +\bar{p}$ events, with $M(e^+e^-)
      >$ 10 GeV/c$^2$ (up to 38 GeV/c$^2$), the QED $\gamma\gamma \rightarrow e^+e^-$ process~\cite{lpair}. 
      Exclusive 2-photon processes had not previously been observed in hadron-hadron collisions; 
      the cross section agrees with the
      precise theory prediction. This process has been suggested as a means of calibrating the LHC luminosity; then it must be done
      in the presence of pile-up, and one will need to know the acceptance etc. at the few \% level. More
      interesting for FP420 is that measurement of an exclusive lepton pair gives both forward proton momenta, with
      a precision dominated by the incoming beam momentum spread ($\frac{\delta p}{p}\approx 10^{-4}$, or 700 MeV). One can do this with pile-up,
      selecting dileptons with no associated tracks on the $l^+l^-$ vertex and $\Delta\phi \approx \pi$. One can
      also cut on $p_T(l^+l^-)$ (correlated with $\Delta\phi$), but $\Delta\phi$ has better resolution. In CDF we found
      that a cut $\pi - \Delta\phi < \frac{0.8 GeV}{M(l^+l^-)}$ rads is suitable for QED-produced pairs. For each pair
      one can predict $\xi_1$ and $\xi_2$, and, if a proton is in the FP420 acceptance, compare $\xi_i$ and
      $\xi_{420}$. This can also possibly map the acceptance
      A($\xi,t\approx$0), as the cross section shape is known from QED, and the (Coulomb) protons have very small $t$.

     \begin{figure}[htp]
     \begin{center}
    \includegraphics[width=0.55\textwidth]{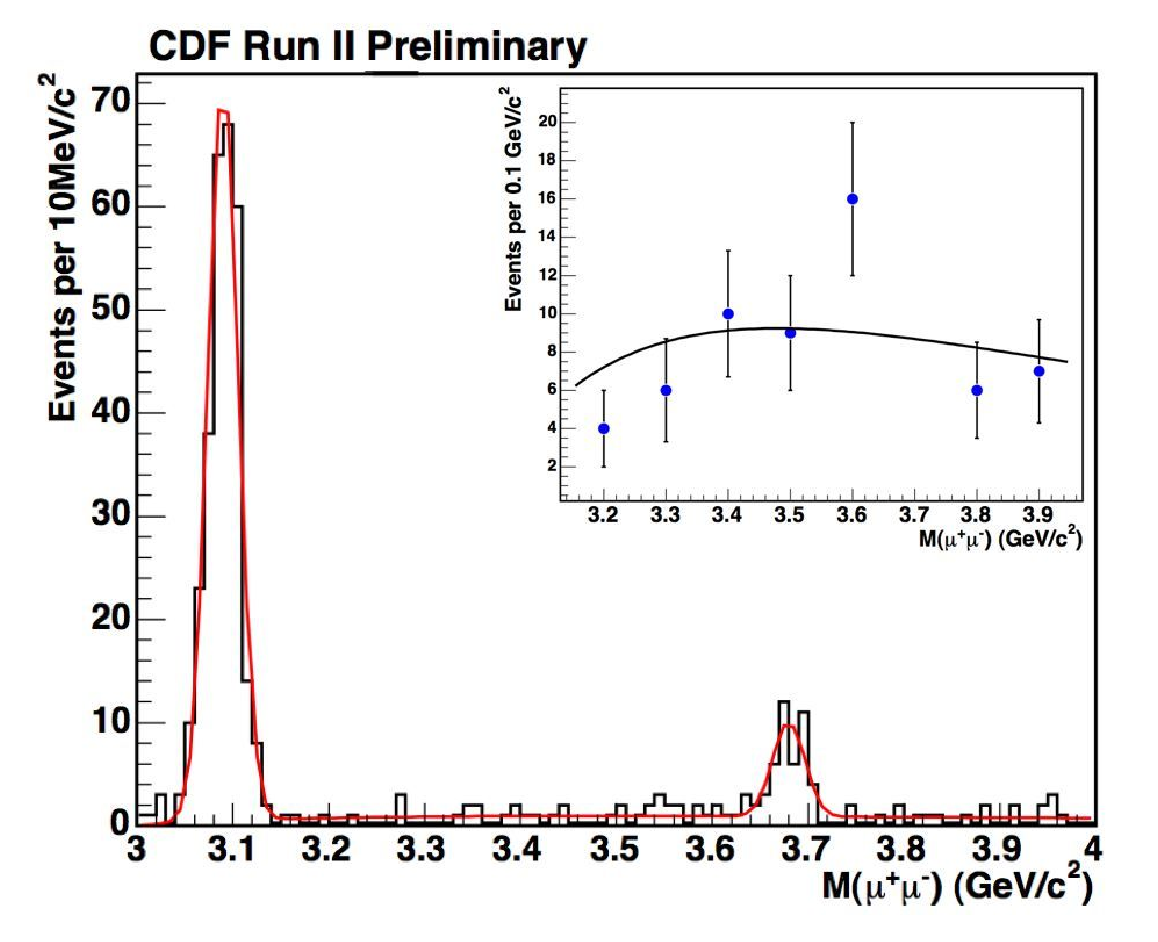}
  \caption{Exclusive dimuon mass spectrum in the charmonium region, together with the sum of two Gaussians and the QED continuum,
  shown in the inset, excluding the 3.65 - 3.75 GeV/c$^2$ bin ($\psi(2S)$). All line shapes are predetermined, with the normalization free.
\label{fig:massmumu} }
\end{center}
\end{figure}

      CDF also used a ``muon+track'' trigger, again with BSC veto, to study $p + \bar{p} \rightarrow p + \mu^+\mu^- +\bar{p}$
      with 3 GeV/c$^2 < M(\mu\mu) < $4 GeV/c$^2$. This is a very rich region, with the $J/\psi$ and $\psi(2S)$ vector mesons
      that can only be produced exclusively by photoproduction $\gamma+I\!\!P \rightarrow \psi$, or possibly by odderon 
      exchange: $O + I\!\!P \rightarrow \psi$. We know what to
      expect for photoproduction from HERA, so an excess would be evidence for the elusive $O$. The spectrum~\cite{mumu34} is
      shown in Fig.~\ref{fig:massmumu}, together with the sum of three components: the vector mesons and a continuum,
      $\gamma\gamma\rightarrow \mu^+\mu^-$, which is again consistent with QED. These central exclusive spectra are exceptionally clean; in fact the biggest
      background ($\approx$ 10\%) is the identical process but with an undetected $p \rightarrow p^*$ dissociation. The
      $J/\psi$ and $\psi(2S)$ cross sections $\frac{d\sigma}{dy}|_{y=0}$, are (3.92$\pm$0.62)nb and (0.54$\pm$0.15)nb, 
      agreeing with expectations~\cite{klein,motyka}. Thus we do not have evidence for $O$
      exchange, and put a limit $\frac{O}{\gamma}< 0.34$ (95\% c.l.), compared with a theory prediction~\cite{bzdak} 0.3 - 0.6.
      
\begin{figure}[htp]
     \begin{center}
    \includegraphics[width=0.50\textwidth]{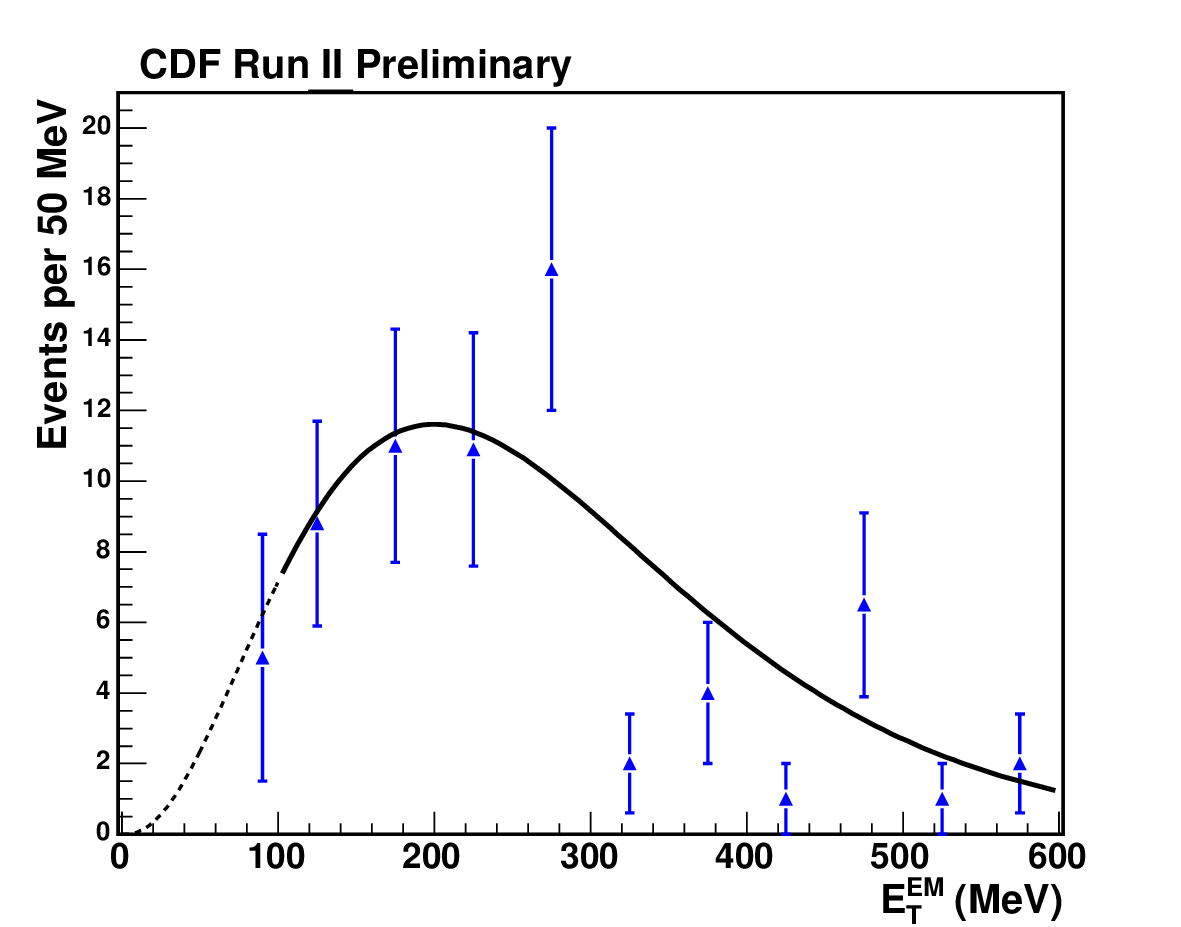}
  \caption{The  $E_T$ spectrum of electromagnetic showers associated with $J/\psi$, together with an empirical
  function to estimate the fraction under the 80 MeV cut. These are $\chi_{c0}(3415)$ candidates.
\label{fig:photon} }
\end{center}
\end{figure}

      While the QED and photoproduction processes in Fig.~\ref{fig:massmumu} should hold no surprises, their agreement with
      expectations validates the analysis. We required no $EM$
      tower with $E_T^{EM} >$ 80 MeV. If we allow such signals (essentially $\gamma$'s) the number of $J/\psi$
      events jumps from 286 to 352, while the number of $\psi(2S)$ only increases from 39 to 40. The spectrum of EM showers is shown in
      Fig.~\ref{fig:photon}. These extra $J/\psi$ events are very
      consistent with being $\chi_{c0}(3415) \rightarrow J/\psi+\gamma$, from $I\!\!P I\!\!P \rightarrow \chi_c$, with about
      20\% of the $\gamma$ being not detected (giving a background of 4\% under the exclusive $J/\psi$). We measure
      $\frac{d\sigma}{dy}(\chi_c)|_{y=0}$ = (75$\pm$14)nb. The existence of this process implies that $p+H+p$ must happen
      at the LHC (assuming $H$ exists), as the QCD physics is qualitatively identical. The $\chi_c$ cross section agrees with predictions: 150nb~\cite{yuan} and 
      130$^{\times 4}_{\div 4}$nb~\cite{khoze}. It is therefore likely that $\sigma(p+p \rightarrow p+SMH+p)$ is of order 0.5-5 fb,
      within reach of FP420. In SUSY models the cross section can be much higher~\cite{fp420}.

      We are looking for $p+\bar{p}\rightarrow p + \Upsilon + \bar{p}$ (by photoproduction, or by $O+I\!\!P$), and $I\!\!P+I\!\!P \rightarrow \chi_b$.  The $\Upsilon$ should
      be measurable in the presence of pile-up using $n_{ass}=0$, $\Delta\phi$ 
      and $p_T$ cuts ($n_{ass}$ is the number of additional tracks on the dilepton vertex). We have candidate events, with the $\Upsilon(1S),(2S)$ and $(3S)$ states resolved; cross sections are
      now being determined.  
      The $\chi_b\rightarrow\Upsilon+\gamma$ probably can not be studied in the presence of pile-up, and it is challenging.  
      We have also made a search~\cite{exclz} for exclusive $Z$, allowed only through photoproduction: $\gamma+I\!\!P \rightarrow Z$. 
      In the Standard Model the (integrated) cross section at the Tevatron is too small to see, $\sigma_{excl}(Z)$ = 0.3fb~\cite{motyka} or
      1.3fb~\cite{goncalves}, before branching fractions. In White's pomeron theory~\cite{white} the cross section
 is expected to be much larger, but a quantitative prediction is lacking. Our search uses both $e^+e^-$ and $\mu^+\mu^-$ pairs
 with $M(l^+l^-) >$ 40 GeV/c$^2$. There are 8 exclusive candidates with $\sigma(p+\bar{p} \rightarrow p +(\gamma\gamma
 \rightarrow l^+l^-) + \bar{p}) = 0.24^{+0.13}_{-0.10}$ pb (for $|\eta(\mu)| <$ 4.0), agreeing with $\sigma$(QED) = 0.256 pb. All the events have
 $\pi - \Delta\phi <$ 0.013(rad) and $p_T(\mu^+\mu^-) <$ 1.2 GeV/c. Only one event had a
 $\bar{p}$ in the acceptance of the Roman pots when they were operational, and a track was observed, showing that the
 event was exclusive, and that at the LHC such $l^+l^- + p$ events will be available for calibration. If we remove the
 requirement that the BSC should be empty there are 4 additional events, interpreted as $p\rightarrow p^*$ dissocation. One of
 them has $M(\mu^+\mu^-)\approx M(Z)$ and a larger $\Delta\phi$ and $p_T$ than the others, but we cannot claim it to be
 truly exclusive. We put a limit on exclusive $\sigma_{excl}(Z) <$ 0.96 pb at 95\% c.l. 
 Clearly it will be interesting to look for exclusive $p+Z+p$ at the LHC. 
    In early running of the LHC, when bunch crossings without pile-up are not yet rare, it is important to measure
    these exclusive processes, to the extent possible without complete forward coverage. In CMS we have plans to
    add forward shower counters~\cite{fsc} around the beam pipe to help tag rapidity gaps, together with the ZDC and forward
    hadron calorimeters. With large forward gaps in both directions, a trigger on
    two EM showers with $E_T >$ 4 GeV should be possible, hopefully observing $\Upsilon \rightarrow e^+e^-,
    \gamma\gamma\rightarrow e^+e^-$, $I\!\!PI\!\!P\rightarrow \gamma\gamma$, and $\chi_b \rightarrow \Upsilon+\gamma \rightarrow
    e^+e^-\gamma$. Clean single interactions are surely needed
   needed for the $\chi_b$ and $I\!\!PI\!\!P\rightarrow \gamma\gamma$; both channels are excellent tests of
   $p+H+p$. One may even hope that when exclusive Higgs production is measured, the coupling $ggH$ can be derived
   by comparing the three cross sections!


\begin{thebibliography}{9}
\bibitem{ewerz} See e.g. C.Ewerz, The odderon in Quantum Chromodynamics, hep-ph/0306137 (2003), and references therein.
\bibitem{afs} T.Akesson \emph{et al.} (AFS Collaboration), Nucl.Phys. \textbf{B264}, 154 (1986).
\bibitem{kiev} At this meeting such a study was initiated by the Kiev group (V.Aushev, L.Jenkovszky et al.).
\bibitem{fp420} M.G.Albrow \emph{et al.}, The FP420 R\&D project, Higgs and new physics with forward protons at the LHC,
arXiv:0806.0302 [hep-ex].
\bibitem{albrow} M.G.Albrow and A.Rostovtsev, Searching for the Higgs at hadron colliders using the missing mass method,
hep-ph/0009336.
\bibitem{khoze} V.A.Khoze \emph{et al.}, Eur.Phys.J. C\textbf{35}, 211 (2004); V.A.Khoze, A.D.Martin and M.G.Ryskin, Eur.Phys.J.
C\textbf{14}, 525 (2000); A.De Roeck \emph{et al.}, Eur.Phys.J. \textbf{C25}, 391 (2002).
\bibitem{exclgg} A.Abulencia \emph{et al.}, (CDF Collaboration), Phys.Rev.Lett. \textbf{99}, 242001 (2007).
\bibitem{mumu34} T.Aaltonen \emph{et al.}, (CDF Collaboration) Observation of exclusive charmonium production
and $\gamma\gamma \rightarrow \mu^+\mu^-$ in $p\bar{p}$ collisions at $\sqrt{s}$ = 1.96 TeV; paper in preparation.
\bibitem{goulianos} T.Aaltonen \emph{et al.}, (CDF Collaboration) Phys.Rev. \textbf{D77}, 052004 (2008).
\bibitem{durhgg} V.A.Khoze \emph{et al.}, Eur.Phys.J. C\textbf{38}, 475 (2005).
\bibitem{exclee} A.Abulencia \emph{et al.} (CDF Collaboration), Phys.Rev.Lett. \textbf{98}, 112001 (2007).
\bibitem{lpair} J.Vermaseren, \textsc{lpair}, Nucl.Phys. \textbf{B229}, 347 (1983).
\bibitem{klein} E.g. S.Klein and J.Nystrand, Phys.Rev.Lett. \textbf{92}, 142003 (2004).
\bibitem{motyka} L.Motyka and G.Watt, Phys.Rev. \textbf{D78}, 014023 (2008).
\bibitem{bzdak} A.Bzdak \emph{et al.}, hep-ph/07021354 (2007). 
\bibitem{yuan}  F.Yuan, Phys.Lett. \textbf{B510}, 155 (2001).
\bibitem{exclz} CDF Collaboration, Search for exclusive $Z$ boson production; paper in preparation.
\bibitem{goncalves} V.P.Goncalves and M.V.T.Machado, Eur.Phys.J. \textbf{C53}, 33 (2008)
\bibitem{white} A.R.White, Phys.Rev. \textbf{D72}, 036007 (2005). White does not claim that 
photoproduced $Z$ have to be exclusive. 
\bibitem{fsc} M.Albrow \emph{et al.}, Forward physics with rapidity gaps at the LHC, arXiv:0811:0120[hep-ex].

\end{thebibliography}
\end{document}